\documentclass[pre,twocolumn,showpacs]{revtex4}

\def\be{\begin{equation}}
\def\de{\end{equation}}
\def\e{\epsilon}
\def\f{\frac}
\def\me{{\bf M}_d\cdot {\bf e}}

\usepackage{times}
\usepackage{epsf}
\usepackage{graphicx}

\begin{document}
\title{Nonlinear alternating current responses of dipolar fluids}
\author{J.\,P. Huang}
\affiliation{Max Planck Institute for Polymer Research, Ackermannweg 10, D-55128, Mainz, Germany, and
Department of Physics, The Chinese University of Hong Kong, Shatin, New Territories, Hong Kong}
\author{K.\,W. Yu}
\affiliation{Department of Physics, The Chinese University of Hong Kong, 
Shatin, New Territories, Hong Kong}
\author{Mikko Karttunen}
\affiliation{Biophysics and Statistical Mechanics Group,
Laboratory of Computational Engineering,
Helsinki University of Technology,
P.\,O.~Box 9203, FIN-02015 HUT, Finland}

\date {\today}

\begin{abstract}

The frequency-dependent nonlinear dielectric increment of dipolar fluids in  
nonpolar fluids is often measured by using a stationary relaxation method in which
two electric fields are used: The static direct current (DC) field of high 
strength causing the dielectric nonlinearity, and the probing alternating 
current (AC) field of low strength and high frequency. When a nonlinear 
composite is subjected to a sinusoidal electric field, the electric 
response in the composite will, in general, consist of AC fields at 
frequencies of higher-order harmonics. Based on the Fr\"{o}hlich model, 
we present a theory to investigate nonlinear 
AC responses of dipolar fluids containing both polarizable monomers 
and dimers. In the case of monomers only, our theory reproduces the 
known results. We obtain the fundamental, second-, and 
third-order harmonics of the Fr\"{o}hlich field by performing a 
perturbation expansion. The even-order harmonics are
induced by the coupling between the AC and DC fields although 
the system under consideration has a cubic nonlinearity only. 
The harmonics of the Fr\"{o}hlich field can
be affected by the field frequency, temperature, dispersion 
strength and the characteristic frequency of the dipolar fluid, as well 
as the dielectric constant of the nonpolar fluid. The results are found 
to be in agreement with recent experimental observations.

\end{abstract}


\maketitle

\section{Introduction}

Electric fields of high strength applied to dipolar fluids produce a nonlinearity 
in the dependence of the polarization on the field strength. On the other hand,
when molecules are placed in an alternating electric field, they display relaxation 
behavior -- for small molecules that occurs typically at microwave frequencies.   
As studying the permittivity of molecules, or larger entities such as cells, 
yields detailed information about the structure and interactions in the system, 
frequency-dependent nonlinear dielectric increment of dipolar fluids has recently
received much attention, see e.\,g. Refs.~\cite{PK,JJ,JJb,PKb} and references 
therein.

Nonlinear dielectric spectroscopy is a useful method for
detecting intermolecular interactions and studying their kinetics
due to its exceptional sensitivity to a large range 
of different aggregation processes in fluids. This method
has been used, for example, to investigate the kinetics of cis-lactams 
cyclic dimerization~\cite{LH,KD}, cholesterol aggregation~\cite{JJc} 
in nonpolar media, and it has been used to 
measure the nonlinear dielectric increment by using a 
stationary relaxation method~\cite{PK,JJ,JJb,PKb}.
Other recent and exciting developments in  dielectric spectroscopy
include, e.\,g.,  studies of phase transitions in liquid crystals~\cite{orihara},
the dielectric properties of 
living cells using high-T$_\mathrm{c}$ SQUIDS at low frequencies~\cite{prodanc},
and combining it with microfluidic devices~\cite{facer} to analyze cells,
proteins and bacteria noninvasively and without chemically modifying the system.

The nonlinear dielectric effect in dipolar fluids is not a result of a single 
mechanism, but it is caused by several factors: 
(i) the Langevin saturation of the orientation of the dipole due to a strong field, 
(ii) intermolecular interactions (e.\,g., dipole-dipole and hydrogen bonds) 
which lead to formation of aggregates with a compensated dipole moment, e.\,g., 
two monomers may form a dimer, $A+B \to  AB$, and (iii) intramolecular 
processes may produce new dipoles, e.\,g., one dimer may dissociate to 
two monomers, $AB \to A+B$. 

The first Langevin effect decreases the permittivity of the fluid, 
whereas the other two effects increase it.
In fact, for aggregation and dissociation processes [(ii) and (iii) above] 
there is a dynamic equilibrium which may depend on the external field. 
Moreover, in the case of a field, additional dimers have to be 
divided to reach a new equilibrium due to the activation energy of the field.

When a nonlinear composite is subjected to a sinusoidal electric field, 
the electric response in the composite will, in general, consist of alternating 
current (AC) fields at frequencies of higher-order 
harmonics~\cite{Levy,Hui,Kling,Gu2,PRE1,JAP1,PRE6}. 
In three previous papers~\cite{PRE1,JAP1,PRE6}, we studied the nonlinear AC 
responses of colloidal suspensions like electrorheological fluids etc. 

In this paper, we present a theory to investigate the 
nonlinear AC responses of dipolar fluids. We show how the harmonics of 
the Fr\"{o}hlich field can be affected by the field frequency, temperature, 
dispersion strength and the characteristic frequency of the dipolar fluid, 
as well as the dielectric constant of the nonpolar fluid in question.  
Thus, by measuring 
the nonlinear AC responses of dipolar fluids, it is possible to investigate 
the frequency-dependent nonlinear dielectric increment.

This paper is organized as follows. In Sec.~\ref{sec:forma}, we derive 
the expression for the nonlinear dielectric increment of a dipolar fluid 
consisting of polarizable dimers and monomers, and obtain the desired fundamental, 
second-, and third-order harmonics of the Fr\"ohlich field. In Sec.~\ref{sec:numer},
we numerically calculate the harmonics of the Fr\"ohlich field  for a 
simple case in which there are no correlations between the orientations
of the molecules. This is followed by a discussion and conclusion in 
Sec.~\ref{sec:disc}.

\section{Formalism\label{sec:forma}}

Let us consider a system in which polar molecules (dimers and monomers) 
are dissolved in a nonpolar fluid of dielectric constant $\e_b$.
If one applies an external field, a dimer may dissociate to two monomers.
We denote the dipole moment of a monomer by ${\bf \mu}$, that of a dimer by 
${\bf \mu}'$, and assume that $\mu'\ll\mu$. In addition, we assume that
only monomers and dimers 
exist (i.\,e., there are no multimers present) and that all dimers have the same dipole moment. 
Furthermore, we assume that the time chemical reactions need is 
short enough to be neglected, and that both dimers 
and monomers have an isotropic polarizability.

\subsection{Nonlinear dielectric effect}

At a high field intensity the dependence of the dielectric displacement ${\bf D}$ 
on the field strength becomes nonlinear~\cite{Bot,DS} 
\be
{\bf D}=\e_0 {\bf E}+4 \pi \chi E^2{\bf E}\label{DE2},
\de
where ${\bf E}$ is the electric field, $\e_0$ the field-independent part of the permittivity,
and $\chi$ is the (nonlinear) susceptibility. To determine  ${\bf D}$, one resorts to 
Clausius-Mossotti equation for three components
\begin{eqnarray}
\f{\e_0-\e_b}{\e_0+2\e_b}&=&\f{4\pi}{3}\left[\eta 
N_1 \left(\alpha_1+\f{\mu'^2}{3k_\mathrm{B}T}\f{1}{1+i \omega\tau_1}\right) +\nonumber \right.\\
  & & \left. (1-\eta) N_2\left(\alpha_2+\f{\mu^2}{3k_\mathrm{B}T}\f{1}{1+i\omega\tau_2} \right) \right], \nonumber
\end{eqnarray}
where $\e_b$ is the dielectric constant of the nonpolar host fluid,
$\eta=N_1/(N_1+N_2)$ and $i=\sqrt{-1}$.  $N_1$ and $N_2$ are the number densities of dimers and 
monomers, respectively, $k_\mathrm{B}$ the Boltzmann constant,
$T$  the absolute temperature, $\omega$  angular frequency of the applied field, and 
$\tau_1$ ($\tau_2$)  the relaxation time of dimers (monomers).
The polarizabilities of dimers and monomers are denoted by $\alpha_1$ and $\alpha_2$, 
respectively.

The desired field-dependent incremental dielectric constant $\e_E$ is~\cite{Bot}
\be
\e_E=\f{\partial D}{\partial E}=\e_0+12 \pi \chi E^2.
\de
Then, the nonlinear dielectric effect is characterized by the ratio
$\triangle \e/E^2$~\cite{Bot}
\be
\f{\triangle \e}{E^2}=\f{\e_E-\e_0}{E^2}=12\pi \chi.
\label{def}
\de

Next, we consider the orientation polarization $p_{or}$ of a dielectric 
sphere of volume $V$ containing $n_1$ 
dimers and $n_2$ monomers with a dipole moment ${\bf \mu_d}$ (see below) 
embedded in a continuum with dielectric constant
 $\e_{\infty}$ (see below). 
  The sphere is surrounded by an infinite dielectric with the same 
macroscopic properties
  as the sphere. 

The average component of the dipole moment in the direction of the field 
due to the dipoles in the sphere
is given by
\be
\langle {\bf M}_d\cdot {\bf e}\rangle =V p_{or}=\f{\int {\rm d}X^{n_1+n_2}{\bf 
M_d}\cdot {\bf e}e^{-u/k_\mathrm{B}T}}{\int {\rm d}X^{n_1+n_2}e^{-u/k_\mathrm{B}T}},
\label{Md}
\de
where ${\bf e}$ denotes the unit vector in the direction of the external 
field, and $X$ stands for the set of positional and orientational 
variables of all monomers and dimers. Here $u$ is the 
energy related to the dipoles in the sphere, and it consists of four parts:
\begin{enumerate}
\item The energy of the dipoles in the external field. 
\item The electrostatic interaction energy  of the dipoles. 
\item The non-electrostatic interaction energy between the dipoles. This 
interaction is responsible for the short-range orientational and positional correlations.
\item The binding  energy (denoted as $u_B$) between two {\it ``monomer dipoles"} in 
the dimer dipoles. 
\end{enumerate}

In Eq.~(\ref{Md}),
${\bf M}_d$ is given by 
\begin{eqnarray}
{\bf M}_d &=& 
\sum_{i=1}^{n_2}({\bf \mu}_d)_i+\sum_{i=1}^{n_1}({\bf \mu}'_d)_i,\nonumber
\end{eqnarray}
with ${\bf \mu}_d = {\bf \mu}(\e_{\infty}+2\e_b)/3\e_b$, and
${\bf \mu}_d' = {\bf \mu}'(\e_{\infty}+2\e_b)/3\e_b$.
The number of monomers and dimers in the system are denoted by
$n_1$ and $n_2$. Parameter $\e_{\infty}$  is the dielectric constant at frequencies 
at which the permanent dipoles (i.e., the orientational polarization) 
cannot follow the changes of the field, but in which the atomic and the 
electronic polarization are still the same as in the static field.
Therefore, $\e_{\infty}$ is the dielectric constant characteristic for 
the induced polarization. In practice,  $\e_{\infty}$ can be expressed 
using an expression containing an intrinsic dispersion 
\be
\e_{\infty} = \e_{\infty}(0)+\frac{\Delta\e}{1+i\omega/\omega_c},
\de
where $\e_{\infty}(0)$ is the dielectric constant at the high-frequency limit, 
and $\Delta\e$ stands for the dielectric dispersion strength with a characteristic frequency $\omega_c$.

The external field in this model is equal 
to the field within the spherical cavity filled with a continuum of dielectric constant
$\e_{\infty}$, while the cavity is situated in a dielectric with 
dielectric constant $\e_0$. 
This field is called the Fr\"{o}hlich field $E_F$, as given by
the Fr\"ohlich model~\cite{Fro}
\be
E_F=\frac{3\e_0}{2\e_0+\e_{\infty}}E+\frac{12\pi\chi 
\e_{\infty}}{(2\e_0+\e_{\infty})^2}E^3.\label{Field}
\de
In this equation, the higher-order terms have been omitted.

With the above definitions and using $u$ as defined in Eq.~(\ref{Md}), we now have
\be
\frac{\partial u}{\partial E_F}=-{\bf M}_d\cdot {\bf e}+\beta.
\de
Here $\beta=\partial u_B/\partial E_F < 0$, and it is first assumed to be a 
quantity independent of $E_F$.
Then, we obtain
\begin{widetext}
\begin{eqnarray}
\left. \langle {\bf M}_d\cdot {\bf e}\rangle \right|_{E_F=0} & = & 0 \nonumber\\
 \frac{\partial}{\partial_{E_F}}\langle {\bf M}_d\cdot {\bf e}\rangle & =&\frac{1}{k_\mathrm{B}T}
[\langle {\bf M}_d\cdot {\bf e}(\me-\beta)\rangle - \langle \me\rangle \langle \me-\beta\rangle ]\nonumber\\
\left. \frac{\partial}{\partial_{E_F}}\langle {\bf M}_d\cdot {\bf 
e}\rangle  \right|_{E_F=0}&=&\f{1}{k_\mathrm{B}T}\langle M_d^2\rangle _0\nonumber\\
 \f{\partial^3}{\partial E_F^3}\langle \me\rangle & = & \f{1}{(k_\mathrm{B}T)^3}[\langle \me 
(\me-\beta)^3\rangle -  3\langle \me (\me-\beta)^2\rangle \nonumber\\
& &         \langle \me-\beta\rangle +6\langle \me(\me-\beta)\rangle \langle \me-\beta\rangle ^2
         -3\langle \me (\me-\beta)\rangle \langle (\me-\beta)^2\rangle \nonumber \\ 
&& -6\langle \me\rangle 
        \langle \me-\beta\rangle ^3+6\langle \me\rangle \langle \me-\beta\rangle \langle (\me-\beta)^2\rangle 
        -\langle \me\rangle \langle (\me-\beta)^3\rangle ]\nonumber\\
 \left. \f{\partial^3}{\partial 
E_F^3}\langle \me\rangle  \right|_{E_F=0}&=&\f{1}{15(k_\mathrm{B}T)^3}[3\langle M_d^4\rangle _0-5\langle M_d^2\rangle ^2_0].\nonumber
\end{eqnarray}
\end{widetext}
Note the subscript $0$ indicates the absence of the field.

To express the macroscopic saturation behavior in microscopic quantities, 
higher derivatives of the average dipole moment have to be taken into account:
\begin{eqnarray}
\langle \me\rangle &=& \left. \f{\partial\langle \me\rangle }{\partial 
E_F} \right|_{E_F=0}E_F+\f{1}{6}\left. \f{\partial^3\langle \me\rangle }{\partial 
E_F^3} \right|_{E_F=0}E_F^3\nonumber\\
&=&\frac{\e_0}{2\e_0+\e_{\infty}}\f{\langle M_d^2\rangle_0}{k_\mathrm{B}T}E+\f{4\pi \chi 
\e_{\infty}}{(2\e_0+\e_{\infty})^2}\f{\langle M_d^2\rangle _0}{k_\mathrm{B}T}E^3\nonumber\\
& 
&+\f{27\e_0^3}{(2\e_0+\e_{\infty})^3}\f{3\langle M_d^4\rangle _0-5\langle M_d^2\rangle _0^2}{90(k_\mathrm{B}T)^3}E^3.\nonumber
\end{eqnarray}
Terms of higher than 3rd order are neglected.

In addition, we have a general relation
\be
\langle \me\rangle =Vp_{or}=\f{\e_0-\e_{\infty}}{4\pi}VE+\chi VE^3.
\de
Using Eq.~(\ref{def}) and the terms in $E^3$ and $E$, we obtain
\begin{eqnarray}
\f{\triangle 
  \e}{E^2}& = &\frac{18\pi}{5(k_\mathrm{B}T)^3}\f{\e_0^4}{(2\e_0+\e_{\infty})^2(2\e_0^2+\e_{\infty}^2)} \times \nonumber\\
&\,& \times [3(\langle M_d^4\rangle _0/V)-5(\langle M_d^2\rangle _0^2/V)].
\label{DelE}
\end{eqnarray}

On the other hand, we may write
\begin{eqnarray}
\f{\langle M_d^2\rangle _0}{V}&=&\left(\f{\e_{\infty}+2\e_b}{3\e_b}\right)^2 [ N_2 
\mu^2\sum_{j=1}^{n_2}\langle \cos \theta_{ij}\rangle  \nonumber\\
                & &+N_1 \mu'^2\sum_{m=1}^{n_1}\langle \cos \theta_{lm}\rangle \nonumber\\
                & &+N_1 \mu \mu'\sum_{q=1,p\neq q}^{n_1}\langle \cos 
\theta_{pq}\rangle \nonumber\\
                & &+N_2 \mu \mu'\sum_{h=1,g\neq h}^{n_2}\langle \cos 
\theta_{gh}\rangle ],\nonumber\\
\f{\langle M_d^4\rangle _0}{V}&=&\left(\f{\e_{\infty}+2\e_b}{3\e_b}\right)^4[N_2 
\mu^4\sum_{j=1}^{n_2}\langle \cos 
\theta_{ij}\sum_{r=1}^{n_2}\sum_{s=1}^{n_2}\cos \theta_{rs}\rangle \nonumber\\
                & &+N_1 \mu'^4\sum_{m=1}^{n_1}\langle \cos 
\theta_{lm}\sum_{u=1}^{n_1}\sum_{v=1}^{n_1}\cos \theta_{uv}\rangle \nonumber\\
                & &+N_1\mu^2\mu'^2\sum_{q=1,p\neq q}^{n_1}\langle \cos 
\theta_{pq}\sum_{c=1}^{n_1}\sum_{d=1}^{n_1}\cos \theta_{cd}\rangle \nonumber\\
                & &+N_2\mu^2\mu'^2\sum_{h=1,g\neq h}^{n_2}\langle \cos 
\theta_{gh}\sum_{a=1}^{n_2}\sum_{b=1}^{n_2}\cos \theta_{ab}\rangle ].\nonumber
 \end{eqnarray}
Hence the nonlinear dielectric increment $\triangle \e/E^2$ is 
explicitly expressed. By setting $N_1=0$, $\mu'=0$ and $\e_b=1$ (no 
dimers in the system), 
Eq.~(\ref{DelE}) can be reduced to Eq.~(7.40) of Ref.~\cite{Bot} where a polar fluid 
in vacuum is investigated. 

It is worth noting that the expression for 
$\triangle \e/E^2$ [Eq.~(\ref{DelE})] contains no $\beta$, i.e., the binding energy 
has no effect on the nonlinear dielectric increment. 
In fact, even if $\beta=\beta(E_F)$, the same result is obtained 
due to the fact that the binding energy is not due to hydrogen bonds, 
rather than from dipoles. 

Let us choose a simple case for numerical calculations. It is assumed that 
there are no correlations between the orientations of the molecules. Thus, we have
\begin{eqnarray}
\sum_{j=1}^{n_2}\langle \cos \theta_{ij}\rangle &=&1\nonumber\\
\sum_{j=1}^{n_2}\langle \cos \theta_{ij}\sum_{r=1}^{n_2}\sum_{s=1}^{n_2}\cos 
\theta_{rs}\rangle &=&\f{1}{3}(5 n_2-2)\nonumber\\
\sum_{m=1}^{n_1}\langle \cos \theta_{lm}\rangle &=&1\nonumber\\
\sum_{m=1}^{n_1}\langle \cos \theta_{lm}\sum_{u=1}^{n_1}\sum_{v=1}^{n_1}\cos 
\theta_{uv}\rangle &=&\f{1}{3}(5 n_1-2)\nonumber\\
\sum_{q=1,p\neq q}^{n_1}\langle \cos \theta_{pq}\rangle &=&0\nonumber\\
\sum_{q=1,p\neq q}^{n_1}\langle \cos 
\theta_{pq}\sum_{c=1}^{n_1}\sum_{d=1}^{n_1}\cos 
\theta_{cd}\rangle &=&\f{2}{3}(n_1-1)\nonumber\\
\sum_{h=1,g\neq h}^{n_2}\langle \cos \theta_{gh}\rangle &=&0\nonumber\\
\sum_{h=1,g\neq h}^{n_2}\langle \cos 
\theta_{gh}\sum_{a=1}^{n_2}\sum_{b=1}^{n_2}\cos 
\theta_{ab}\rangle &=&\f{2}{3}(n_2-1)\nonumber
\end{eqnarray}
In this regard, the nonlinear dielectric increment is given by
\begin{eqnarray}
\f{\triangle \e}{E^2}
& = &\f{18\pi}{5(k_\mathrm{B}T)^3}\f{\e_0^4}{(2\e_0+\e_{\infty})^2(2\e_0^2+\e_{\infty}^2)}\times \nonumber\\
& \, &\times \left(\f{\e_{\infty}+2\e_b}{3\e_b}\right)^4\phi,
\label{non}
\end{eqnarray}
where
\begin{eqnarray}
\phi &=& \mu^4(-5VN_2^2+5N_2n_2-2N_2)\nonumber\\  
& & +2\mu^2\mu'^2(N_1n_1-N_1+N_2n_2-N_2-5VN_1N_2)\nonumber\\  
& & +\mu'^4(-5VN_1^2+5N_1n_1-2N_1).\nonumber
\end{eqnarray}

\subsection{Nonlinear AC responses}

\begin{figure}
\includegraphics[
width=.35\textwidth]{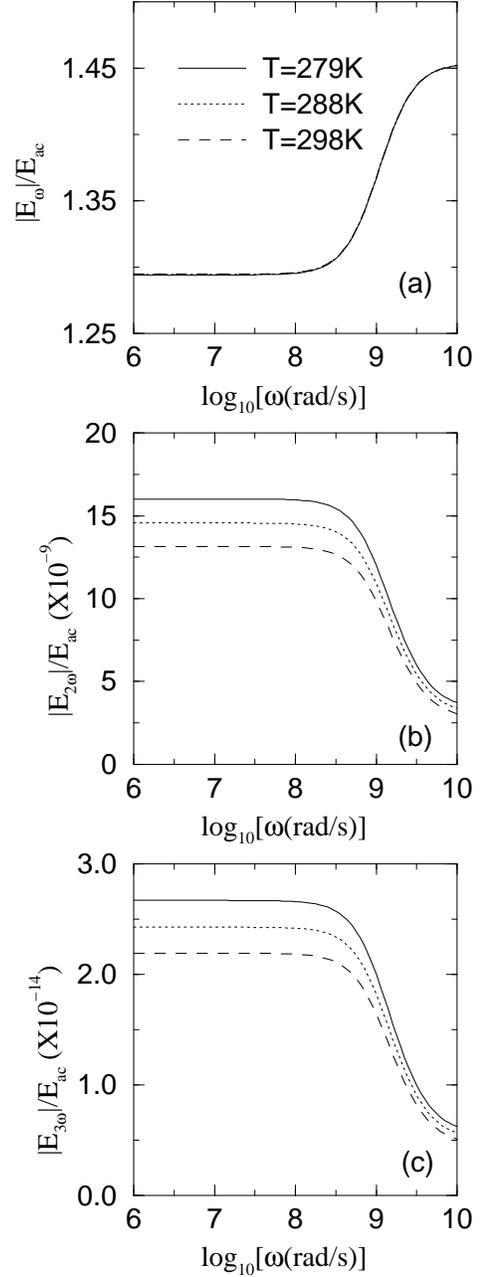}
\caption{The fundamental, second- and third-order harmonics of the Fr\"{o}hlich field 
as a function of the field frequency for three different temperatures.}
\label{fig1}
\end{figure}

Next, we perform a perturbation expansion~\cite{Gu1} to investigate 
the nonlinear AC response of the dipolar fluid. In this case, the applied 
field is ${\bf E} = {\bf E}_\mathrm{dc}+{\bf E}_\mathrm{ac}(\omega)$, 
where ${\bf E}_{dc}$ denotes the DC field, and  ${\bf E}_\mathrm{ac}(\omega)$ 
the sinusoidal AC field, namely 
${\bf E}_\mathrm{ac}(\omega)={\bf E}_\mathrm{ac}\sin (\omega t) .$ 
As a result, the Fr\"{o}hlich 
field is expressed in the terms of harmonics~\cite{Gu2,PRE1,PRE6,Wei} as
\begin{eqnarray}
E_F &= &E_F^{(\mathrm{dc})}+E_{\omega}\sin (\omega t)+ \nonumber \\
&& +E_{2\omega}\cos (2\omega t)+E_{3\omega}\sin(3\omega t) + \cdots,
\label{EfHar}
\end{eqnarray}
where the DC component ($E_F^{\mathrm{(dc)}}$) and the fundamental, second-, 
and third- order harmonics ($E_{\omega}$, $E_{2\omega}$  and $E_{3\omega}$) 
are analytically given by
\begin{eqnarray}
E_F^{(\mathrm{dc})} &=& \frac{3\epsilon_0}{2\e_0+\e_{\infty}}E_\mathrm{dc}
+\frac{12\pi\chi\e_{\infty}}{(2\e_0+\e_{\infty})^2}E_\mathrm{dc}^3 \nonumber \\
& & +\frac{18\pi\chi\e_{\infty}}{(2\e_0+\e_{\infty})^2}E_\mathrm{dc}E_\mathrm{ac}^2,\\
E_{\omega} &=& \frac{3\e_0}{2\e_0+\e_{\infty}}E_{ac}+\frac{36\pi\chi\e_{\infty}}
{(2\e_0+\e_{\infty})^2}E_\mathrm{dc}^2E_\mathrm{ac}\nonumber \\
& & +\frac{9\pi\chi\e_{\infty}}{(2\e_0+\e_{\infty})^2}E_\mathrm{ac}^3,\\
E_{2\omega} &=& -\frac{18\pi\chi\e_{\infty}}{(2\e_0+\e_{\infty})^2}E_\mathrm{dc}E_\mathrm{ac}^2,\\
E_{3\omega} &=& -\frac{3\pi\chi\e_{\infty}}{(2\e_0+\e_{\infty})^2}E_\mathrm{ac}^3.
\end{eqnarray}
In the above derivation, we have used two identities, 
$\sin^2 (\omega t)=[1-\cos (2\omega t)]/2$ and $\sin^3 (\omega t) 
= (3/4)\sin(\omega t) - (1/4) \sin (3\omega t)$. 
It is worth noting that the Fr\"{o}hlich field [Eq.~(\ref{EfHar})] is 
a superposition of both odd- and even-order harmonics, even 
though there is initially a cubic nonlinearity only [see, Eq.~(\ref{DE2})]. 
Actually, the occurrence of both odd- and even-order harmonics 
is simply due to the coupling between the DC and the AC fields~\cite{Wei}. 
In Eq.~(\ref{EfHar}), we have omitted higher-order harmonics 
(e.\,g., fourth-order, fifth-order, and so on).

\section{Numerical Results \label{sec:numer}}

\begin{figure}[tb]
\includegraphics[width=.35\textwidth]{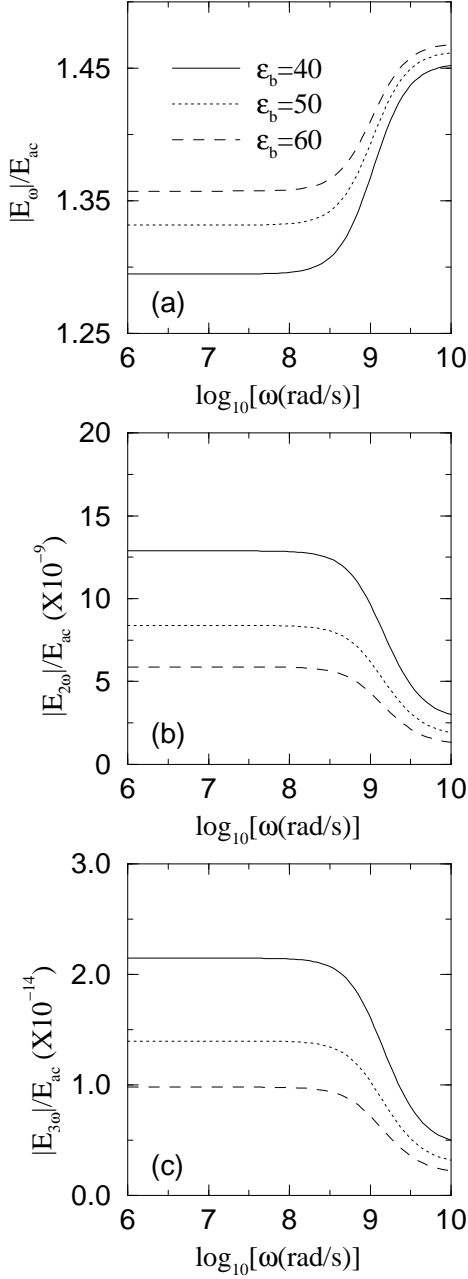}
\caption{Same as Fig.~\ref{fig1}, but for three different $\epsilon_{b}$ 
(dielectric constant of the host fluid ).}
\label{fig2}
\end{figure}

\begin{figure}[tb]
\includegraphics[width=.35\textwidth]{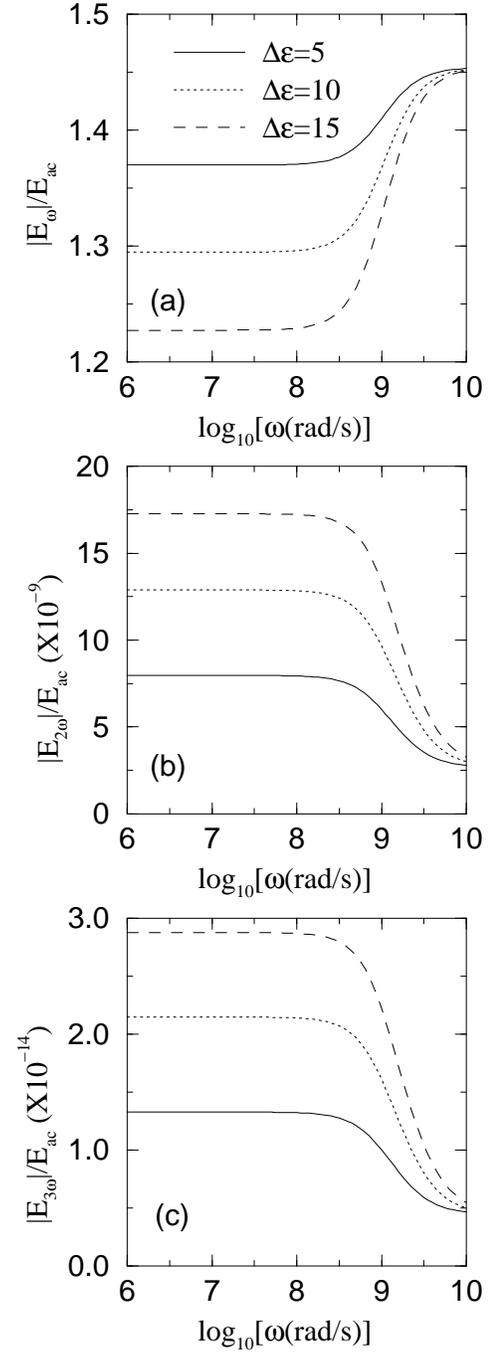}
\caption{Same as Fig.~\ref{fig1}, but for three different $\Delta\epsilon$.  }
\label{fig3}
\end{figure}
\begin{figure}[tb]
\includegraphics[width=.35\textwidth]{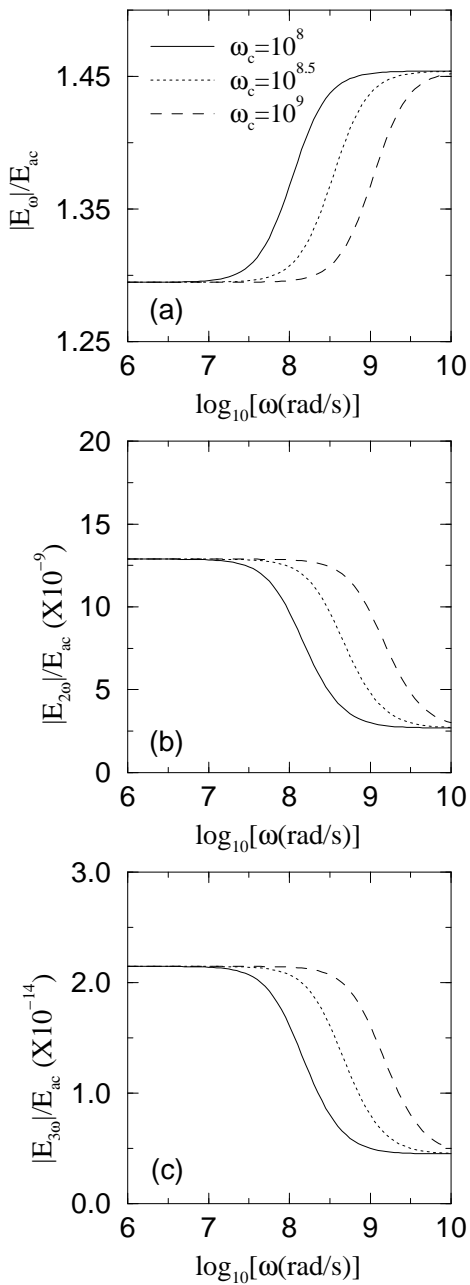}
\caption{Same as Fig.~\ref{fig1}, but for three different $\omega_c$.  }
\label{fig4}
\end{figure}

Based on Eqs.~(\ref{def}),~(\ref{non}),~and~(\ref{EfHar}), 
we are now in a position to perform numerical calculations. 
Without loss of generality, we choose the following parameters: 
$V=5.0\times 10^5\,$cm$^3$, 
$n_1=0.25\times 10^{15}$, $n_2=1.0\times 10^{15}$, $\mu=2.5D$, 
$\mu'=0.05D$, $\tau_1=0.018\,$ns, $\tau_2=0.02\,$ns, 
$\alpha_1=8\times 10^{-25}\,$cm$^3$, $\alpha_2=2\times 10^{-25}\,$cm$^3$, $T=300\,$K, 
$\e_b=40 ,$ $\epsilon_{\infty}(0)=2.5,$ $\Delta\epsilon=10,$ $\omega_c=10^9\,$rad/s,  
$E_\mathrm{ac}=1\,$V/cm, and $E_\mathrm{dc} = 10^5\,$V/cm, 
unless otherwise mentioned.
Here $D$ is Debye unit ($10^{-18}$ e.s.u. of electric moment).

Figure~\ref{fig1} displays the fundamental, second- and third-order harmonics  
of the Fr\"{o}hlich field as a function of the field frequency.  It is 
shown that the harmonics of the Fr\"{o}hlich field are sensitive to the 
frequency of the field. Decreasing the frequency causes the second- and 
third-order (fundamental) harmonics to increase (decrease).  Thus, 
it is possible to investigate the frequency-dependent nonlinear 
dielectric increment ($\Delta\epsilon/E^2$) of dipolar fluids by 
detecting the nonlinear AC responses.  We also find that decreasing 
the temperature $T$ causes the second- and third-order harmonics 
to increase due to the change in the Fr\"{o}hlich field. This result 
is in qualitative agreement with the experimental findings by 
Hellemans {\it et al.}~\cite{JJ}. In their experiment, they measured 
the nonlinear dielectric relaxation spectra for 10-TPEB dissolved 
in benzene at three different temperatures, $279\,$K, $288\,$K, 
and $298\,$K. In addition, it is shown that the temperature has 
no effect on the fundamental harmonics 
(note that the three curves in Fig.~\ref{fig1}(a) overlap).

In Fig.~\ref{fig2}, we investigate the effect of  the  dielectric constant 
of the nonpolar fluid  $\epsilon_{b}$ on the harmonics, and find 
that decreasing $\epsilon_{b}$ leads to increasing second- and 
third- order harmonics, but decreasing fundamental harmonics. 
This is due to the change of the Fr\"{o}hlich field. 
However, decreasing $\Delta\e$ has exactly the opposite effect, 
see Fig.~\ref{fig3}.

Figure~\ref{fig4} shows the effect of the intrinsic characteristic 
frequency $\omega_c$. From this figure, we can conclude 
that the characteristic frequency for the harmonics is 
strongly dependent on the intrinsic characteristic frequency of the dipolar fluid.

\section{Discussions and conclusions \label{sec:disc}}

In the present paper, we have studied a system containing 
polarizable monomers and dimers. 
Our theory can be extended to deal with multimers as well.
Here, we have studied the fundamental, second-, and third-order harmonics. 
The extension to higher-order harmonics is interesting and the present
theory can be extended to include them simply
by keeping the higher-order terms in Eqs.~(\ref{Field})~and~(\ref{EfHar}).

Throughout this paper, we have discussed the case of weak nonlinearity. 
Thus, for extracting the harmonics, a perturbation expansion 
was performed. In the case of strong nonlinearity, this approach is no 
longer valid, and we might need to resort to a self-consistent method~\cite{PRE1,Yu96}.
In addition, the dipolar fluid considered here is seen as dispersive. 
In fact, in real systems, the nonpolar fluid is also dispersive. 
The present theory can be extended for that case as well.

To summarize, we have presented a theory to investigate the nonlinear AC responses 
of dipolar fluids based on the Fr\"{o}hlich model. 
It was shown, that our theory reproduces the known results of a polar fluid in 
vacuum~\cite{Bot} and is consistent with the experimental observations of
Hellemans {\it et al.}~\cite{JJ} for the nonlinear dielectric relaxation 
spectra of 10-TPEB dissolved  in benzene.
For the simple case where there 
are no correlations between the orientations of the molecules, 
we have obtained the harmonics of the Fr\"{o}hlich field analytically 
by  performing a perturbation expansion approach. It has been found 
that the harmonics of the Fr\"{o}hlich field are affected by the field 
frequency, temperature, dispersion strength, and the characteristic
frequency of the dipolar fluid, as well as the dielectric constant 
of the nonpolar host fluid. The results are found to be in agreement 
with experimental observations.  
Thus, by measuring the nonlinear 
AC responses of dipolar fluids, it is possible to investigate 
the frequency-dependent nonlinear dielectric increment.

\begin{acknowledgments}
This work has been supported by the DFG under grant No.~HO\,1108/8-4 (J.\,P.\,H) 
and the Academy of Finland grant No.~00119 (M.\,K.). 
K.\,W.\,Y. acknowledges the financial support from RGC Earmarked Grant.
\end{acknowledgments}

\end{document}